# Coupling of Wind Farms with Nuclear Power Plants

Kareem Al-Ashery[1]   Walid El-Khattam[2]   M. Abd El Rehim Badr[3]

**Summary**

Countries in the Arab region are planning to meet their increasing electricity demands by increasing both nuclear and Renewable Energy (RE) contributions in electricity generation. In the planning phase of siting both new Wind Farms (WFs) and Nuclear Power Plants (NPPs), many benefits and challenges exist. An important aspect taken into consideration during the NPP siting is the existence of ultimate heat sink which is sea water in most cases. That is why most NPPs are sited on sea coasts. On the other hand, during WF siting, the main influential aspect is the existence of good wind resources. Many coastal areas around the world fulfill this requirement for WF siting. Coupling both NPPs and WFs in one site or nearby has many benefits and obstacles as well. This coupling can help in the geographical distribution of WFs along the countries territories by installing WFs at or near to the NPP area. This geographical distribution will make use of the low density population area around NPPs and smoothen the WF's output power injected into the grid.

Due to wind speed fluctuations, the value of the output of WFs' power is less than their installed capacity. Thus, WFs' capacity credit is evaluated to find out the corresponding value of the thermal power plants to be installed in the grid in order to satisfy the required system reliability.

In this paper, the implementation aspects of NPP and WF coupling/adjacency will be discussed in detail. Based on international experience and literature reviews, the benefits and obstacles of this coupling/adjacency are studied and evaluated. Various case studies are carried out to verify the coupling/adjacency concept. The benefits of WF geographical distribution are examined on two candidate sites in Egypt. The WF capacity credit is calculated by implementing the PJM method using actual three-year hourly wind data. The obtained results are evaluated to study their applicability in the Egyptian environment and their applicability for countries in the Gulf region. Finally, both the coupling idea and the capacity credit values can be used to help decision makers in the planning phase as well as in the selection of WT characteristics as discussed in this paper.

[1] Egyptian Atomic Energy Authority, Nuclear Research Center, Cairo, Egypt
[2] Electrical Power and Machines Department, Faculty of Engineering, Ain Shams University, Cairo, Egypt (e-mail: walid_el_khattam@hotmail.com)
[3] Future University, Cairo, Egypt

## 1. Introduction

Climate change has been identified as one of the greatest challenges facing nations, governments, businesses and citizens of the globe. The threats of climate change demand an increase in the share of Renewable Energy (RE) from the total of energy generation. In the meanwhile, there are tremendous efforts to decrease the reliance on fossil fuel energies which opens the venue for increasing the usage of alternative resources such as nuclear energy. Many countries in the Arab region, ex. Egypt and the United Arab Emirates (UAE) are planning to meet increasing electricity demands by increasing both RE and nuclear contributions in electricity generation.

Due to limited fossil fuel energy resources in Egypt and near full utilization of hydro energy, the government is moving to alternative sources. Besides others, the wind energy share is planned to increase, and nuclear energy is planned to enter the country energy mix. In 2008, the Supreme Council of Energy has set a target to have a 20% of the total generated electrical energy from RES by the year 2020. Wind energy shall represent 12% (7200 MW) of the RE strategy, 3000 MW of which should be installed on the Red Sea coast [1]. The Wind Atlas of Egypt confirms the existence of wind energy potential areas around the country especially some areas along the Mediterranean and Red seas coasts, and the west and east of the Nile valley. In the meanwhile, in most cases Nuclear Power Plant (NPP) is sited near sea coasts, this is due to the requirements of ultimate heat sink. The first planned NPP will be in El Dabaa site on the Mediterranean Sea coast. The site can accommodate four units each of 1000-1600 MW installed capacity. The first unit is planned to enter service in 2019 and the last in 2025. Searching for sites that can accommodate both nuclear power plants and Wind Farms (WFs), coastal areas in Egypt are appealing for both.

In UAE, targets were set to promote RE as follows; Abu Dhabi 7% (1500 MW) by 2020, and Dubai 5% (1000 MW) by 2030. A wind energy project of 28.8 MW was approved [2]. On the other hand in April 2008, the UAE nuclear energy policy was published. In 2009, the UAE Federal Nuclear Law was adopted, the Emirates Nuclear Energy Corporation was launched, and a prime contractor was selected. Four units, 1400 MW each, will be completed, one each year starting 2017 [3].

Internationally, WF and NPP adjacency projects exist in many countries such as:
- In Illinois (USA) where Grand Ridge WF of capacity 98MW is located adjacent to LaSalle NPP of a capacity of 2309MW [4].
- In Ontario (Canada), the Ripley Wind Power Project of capacity 76MW is located near Bruce Power NPP of capacity 4820MW, with some kilometers between the two sites [5].
- In Ontario (Canada), a single WT of capacity 1.8 MW is erected in the site of Pickering Nuclear Generating Station of capacity 3100 MW for research purposes.
- In Finland, Olkiluoto WF of capacity 1 MW is located near Olkiluoto NPP of capacity 1760 MW.

## 2. Coupling WF and NPP

Literatures reported the pros and cons of coupling WFs and NPP. Some main features are stated and discussed as follows:

➤ Pros of WF and NPP adjacency:

A. *Good use of low population density areas around NPPs*

Multiple planning zones are required around an NPP to minimize health and safety risks:
- Zone-1: exclusion area, the site of the NPP extends to approximately 1 km from the facility. Within this area, permanent settlement is prohibited and the operator of the facility

- should have authority over all activities carried out in the area.
- Zone-2: extends to approximately 5 km from the facility. Development is restricted in this zone to exclude sensitive activities (hospitals,etc.) and high density settlements.
- Zone-3: consists of two sub-zones, a plume exposure and ingestion pathway zones that have radii around 16 km and 80 km respectively [6].

These areas around NPP are very good for WF siting based on wind availability.

*B. Existence of connection to the grid*

WFs have to be installed in the immediate proximity of wind. The best conditions for installation of WFs are usually in remote, open areas with low population densities [7] and [8], and low power transmission capacity [9] which are poorly suited to accommodate WFs power [10].

By integrating WF near NPP, new electrical connection for WF grid integration is avoided. In an NPP site there are at least two substations with different voltage levels (ex. 500KV & 220 KV) and a strong connection to the grid due to the large generation capacity of NPP.

*C. Existence of main facilities required for WFs*

An NPP site contains facilities that are useful for WF during different periods of construction and operation such as strong road network, workers and campus, etc.

*D. Coupling helps in geographical distribution of WFs*

The wind speed varies continuously as a function of time and height. Therefore, wind energy is characterized by large variations in production [7]. Aggregation of WFs output due to geographic distribution dramatically reduces the wind fluctuations and there will be fewer instances of near zero or peak output [7] and [11]. Grid system operators manage such variations using spinning reserves [8]. By geographical distribution of WF, wind power variation is eliminated and thus the extra cost of running more thermal plants as spinning reserves is lowered [11]. Therefore, WFs can be installed in areas around NPP to help in geographical distribution of $WF_S$ in the grid.

*E. Onsite WF and NPP Emergency Power Systems (EPSs)*

An important characteristic of NPP is that after a nuclear reactor is shut down, it continues to produce a significant amount of heat for an extended period. Hence the reactor cooling systems must continue to operate for several days after a reactor shuts down, to prevent overheating and damage to the reactor core. Therefore, reliable cooling arrangements must be provided, and this requires robust and diverse sources of reliable electrical supply [12]. This supply comes either from the grid (off-site power), or from on-site emergency back-up power, such as batteries, diesel generators, or gas turbines. In case of coupling, WFs, as another type of generation, can increase the reliability and availability of power source in the $EPS_S$.

➤ Cons of Coupling of WF and NPP:

*A. High short circuit power level at Point of Common Coupling of WF*

Connecting a WF to the substation where NPP is connected (high short circuit capacity) will require the short circuit duty of the WF switchgear to be high and very expensive.

*B. Bad impact on human activities relating to NPP operation*

This is due to WFs Noise [8] and Shadow flicker [13] and [14], and blade glint [13].

*C. Interference of $WT_S$ with the communication systems of NPP*

*E. Lower energy production from WF*

The idea of coupling is based mainly on constructing WF around NPP. This may lead to erecting WF in an area with lower wind resources compared to other sites in the country. Thus, it results in lowering the energy production for the same WF capacity. This disadvantage can be balanced by the other technical and economic benefits of the coupling.

*F. Improper wind energy power quality aspects and its impact on NPP auxiliaries.*

## 3. Site Selection for Coupling of WF and NPP

The main objective in site selection for nuclear installations is to protect the public and the environment from the radiological consequences of radioactive releases due to accidents or normal operation [15]. On the other hand, the main requirement during WF site selection is the existence of potential wind resource. WF site selection requires a number of technical, environmental, and planning issues [16] and [17]. The site selection requirements of coupling can be divided to three main groups:

- Group-1: Requirements for both WF and NPP site selection, which can be summarized as:
  1. Proximity to appropriate existing electricity infrastructure and transport infrastructure
  2. Sufficient distance from permanent human activities.
  3. Siting away from ecological areas, and cultural heritage areas.
- Group-2: Requirements for NPP site selection, which can be summarized as:
  1. Requirement of ultimate heat sink for cooling in NPP steam cycle [6];
  2. Siting away from external hazards such as earthquakes, surface faulting, flooding, etc…[15].
- Group-3: Requirements for WF site selection, which can be summarized as:
  1. Need for potential wind resource in the site, i.e. average wind speeds above 6 m/s [16].
  2. Avoidance of electromagnetic interference of $WT_S$ with microwave, television, radar, etc.
  3. Avoidance of $WT_S$ physical obstruction for aviation operations in the site [16] and [17].

*Proposed sites for coupling in Egypt*

Coastal areas in Egypt are appealing for both WF and NPP, and the coupling between them is applicable. According to [18], there are six sites under evaluation for NPP in Egypt. Considering the WF and NPP coupling, Table 1 provides the wind speed at the candidate NPP sites from the Wind Atlas [19] which satisfies the WT site requirements, i.e. values are average wind speeds above 6 m/s [16].

Table 1
Average wind speeds at the NPP candidate sites

| Site | V* (m/s) |
|---|---|
| ElDabaa | 6-7 m/s |
| El Negeila East | 6-7 m/s |
| El Negeila West | 6-7 m/s |
| Hammam Pharoaun | 6-8 m/s |
| South Safaga | 6-7 m/s |
| South Mersa Alam | 5-6 m/s |

* Measurements at 50 m/s above ground level

Among others, El Dabaa site is proposed for the first NPP in the country, with an average wind speed of about 6.5 m/s. In [20], a study was carried out to calculate the cost of generating electricity from 2MW-WTs at El Dabaa. Results show that the cost of electricity generation is competitive with the cost from the Egyptian Electricity Authority. Therefore, El Dabaa can be considered as one of the best proposed sites for WT coupling.

## 4. Geographical Smoothing (Case Study):

In this paper, to investigate the impact of the geographical distribution of WFs in Egypt, two cases are examined:
1. Case-A: Geographical distribution of WFs by their installation on both the shore of Suez Bay and near NPP sites (El Dabaa) as proposed by the paper.
2. Case -B: Locating WFs as it is in the Egyptian RE strategy, which is based on maximizing energy yield of the installed capacity. This means locating them only on the shore of Suez Bay.

According to the Egyptian plan for both WFs and NPPs, and the available wind data, two sites are chosen to conduct this study. They are El Dabaa (North Mediterranean coast), and the Zafarana (Suez Bay- The Red Sea). Only mean hourly wind speed per month was made available to the authors in (m/s) at a height 24.5 m above the ground level from [19]. Therefore, 288 values for each site are used. For El Dabaa site, due to the limitation in data, data for EL Galala site (that is near to El Dabaa) is used, Figure 1. The geographical coordination of each site is illustrated in Table 2.

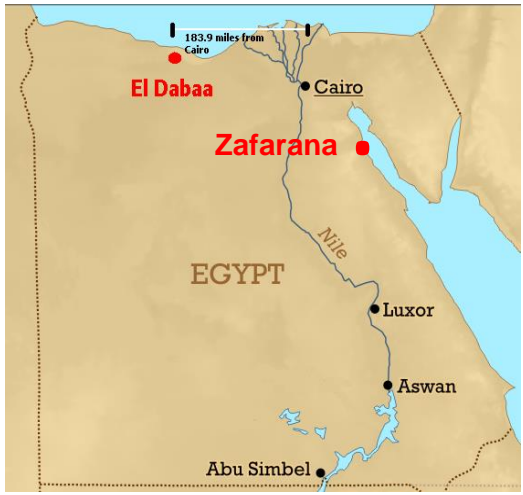

Fig. 1 Map of Egypt

Table 2
Geographical coordination of the candidate sites

| Site | Coordinates | | | | |
|---|---|---|---|---|---|
| | Latitude | | Longitude | | Elevation |
| | Deg. | Min. | Deg. | Min. | (m) |
| El Dabaa | 30 | 56 | 28 | 28 | 17 |
| El Galala | 31 | 1 | 28 | 11 | 59 |
| Zafarana | 29 | 06 | 32 | 36 | 25 |

To calculate the WF generated power, the wind speed has to be measured at the WT hub. Therefore, the actual wind speed at the WT hub (80 m) is calculated as follows:

$$V_T = V \cdot \left(\frac{H_T}{H}\right)^\alpha \quad (1)$$

Where;
V: wind speed in m/s at H height in m,
$V_T$: wind speed at the WT hub height in m,
α: the correction factor that is assumed to be 0.1429 for both sites [21].

The output power of WT can be calculated using the power curve shown in Figure 2 and represented by (2).

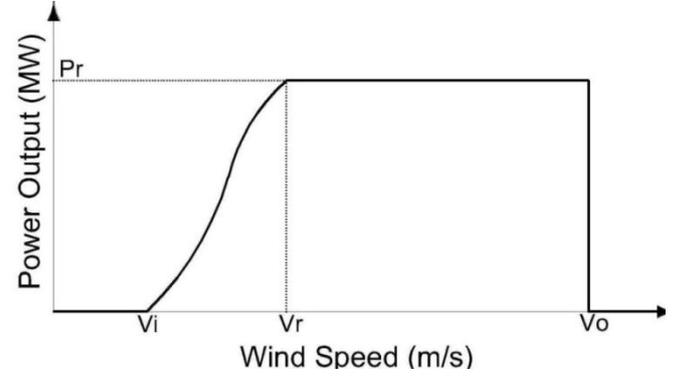

Fig. 2 WT power curve.

$$P(V_t) = \begin{cases} 0 & V_t < V_i \text{ or } V_t > V_o \\ (A + B \times V_t + C \times V_t^2) \times P_r & V_i \leq V_t < V_r \\ P_r & V_r \leq V_t < V_o \end{cases} \quad (2)$$

Where:

$$A = \frac{1}{(V_i - V_r)^2}\left\{V_i(V_i + V_r) - 4V_iV_r\left[\frac{V_i+V_r}{2V_r}\right]^3\right\} \quad (3)$$

$$B = \frac{1}{(V_i - V_r)^2}\left\{4(V_i+V_r)\left[\frac{V_i+V_r}{2V_r}\right]^3 - (3V_i + V_r)\right\} \quad (4)$$

$$C = \frac{1}{(V_i - V_r)^2}\left\{2 - 4\left[\frac{V_i+V_r}{2V_r}\right]^3\right\} \quad (5)$$

Where;
$P_r$: Rated power output of the WT
$V_i$, $V_r$, and $V_o$: Cut-in, rated, and cut-out wind speed of the WT respectively
$V_t$: Wind speed at time (t)
$P(V_t)$: Power output of the WTG at wind speed ($V_t$)

Assuming identical WTs, the total WF power is calculated as follows:

$$P_{WF}(V_t) = N \times P_{WT}(V_t) \quad (6)$$

Where;
$P_{WF}(V_t)$: electrical power generated by WF (MW)

N: Number of WTs in the WF

The power generation in percentage of installed capacity ($\%P_{WF}(V_t)$) for each site will be:

$$P(V_t) = \begin{cases} 0 & V_t < V_i \text{ or } V_t > V_o \\ (A + B \times V_t + C \times V_t^2) & V_i \leq V_t < V_r \\ 1 & V_r \leq V_t < V_o \end{cases} \quad (7)$$

In this paper, a 2 MW-WT type is assumed to be used. The installed WT capacity installed in Zafarana is about 500 MW WF. In addition to what is installed in Zafarana, it is assumed that a new 500 MW to be installed in two candidate sites, Zafarana or El Dabaa to have a total of 1000MW installed in the grid. The power generation for the two WFs can be calculated using the given wind data and (1) to (7).

*Case-A:*
It examined the impact of installing new 500 MW in EL Dabaa to what already exists, 500 MW in Zafarana, which represents dispersed WFs in two different locations.

*Case B:*
A 1000 MW in Zafarana is assumed, which represents two WFs in the same region.

For each case under study (Case-A and Case B), three scenarios are discussed based on the used WT's characteristics. The 500 MW WTs already installed Zafarana site has cut-in, rated, and cut-out wind speeds which are 4, 10, and 23 respectively. The new 500 MW WTs to be installed in Zafarana and El Dabaa have the following characteristics:

*Scenario-I:* Similar to Zafarana WTs.
*Scenario-II:* cut-in, rated, and cut-out wind speeds of 4, 12, and 25 respectively [22].
*Scenario-III:* cut-in, rated, and cut-out wind speeds of 4, 13, and 25 respectively.

Results in Figures 3a-c show that, Case-A has a smoother annual generation in percentage of available capacity than that of Case-B. This is due to the effect of geographical distribution. In the meantime, it can be concluded that as the WT rated power and cut-out speeds increase, the output generated power becomes smother and the impact of geographical distribution is better.

Similar to Figure 3, Figure 4 shows the same conclusions in the duration curves of wind power generation form, i.e. geographical spreading leads to a flattened duration curve for wind power production of Case-A than in Case-B. Furthermore, increasing the WT rated and cut-out speeds, reduces the overall generation level.

To show the variation in output power (in percentage of the installed capacity) from one reading to the next, equation (8) is used:

$$\Delta P_n = P_n - P_{n-1} \quad for \quad 2 \leq n \leq N \quad (8)$$

Where:
$\Delta P_n$: variation in output power (% installed capacity) from one reading to the next
$P_n$: output power (% installed capacity) in according to reading number (n)
n: reading number;
N: total number of readings for each case (24*12=288)

From Figure 5, Table 3 shows that geographical distribution of WFs in Egypt will smoothen the aggregated wind energy production as shown in Case-A. In the mean while, increasing the WT rated and cut-out speeds, reduces the output variations as well. Therefore, smaller output power variations means more efficient planning/utilizing of the cross-section area of the overhead lines going to be used for evacuating the generated power. Thus, leading to economical cost saving.

Table 3
Duration curves of monthly variations, as a percentage of the installed capacity

|        | Scenario-I | Scenario-II | Scenario-III |
|--------|------------|-------------|--------------|
| Case-A | 16 to -18  | 12 to -16   | 11 to - 15   |
| Case-B | 28 to -31  | 24 to -37   | 22 to -34    |

## 5. WT Capacity Credit (Case Study in Egypt)

It is hard to predict the behavior of wind; therefore, the evaluation of capacity credit provided by wind energy must be studied well.

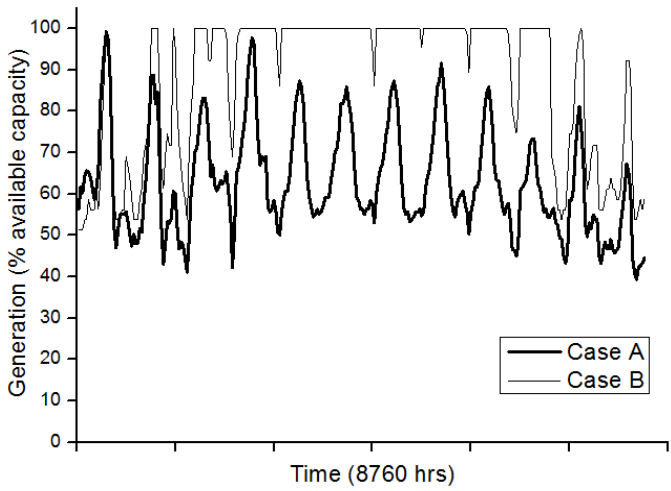

Fig. 3.a Scenario-I

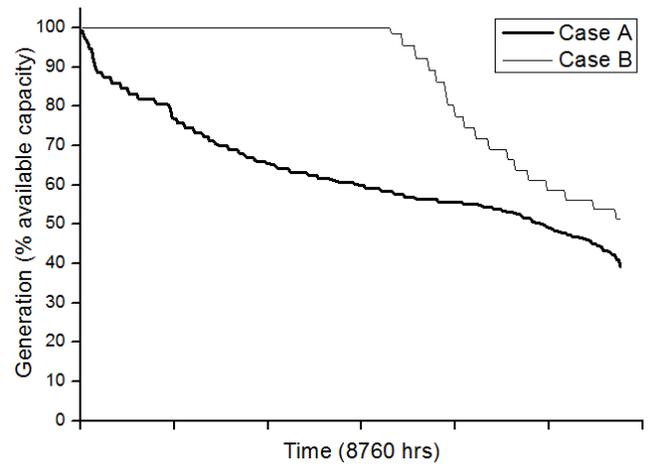

Fig. 4.a Scenario-I

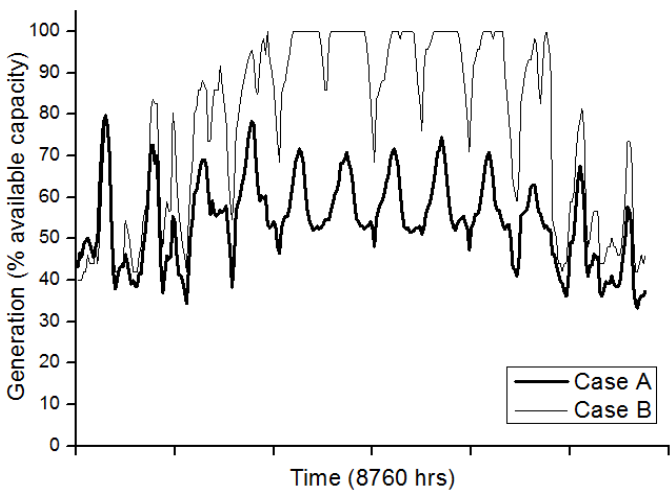

Fig. 3.b Scenario-II

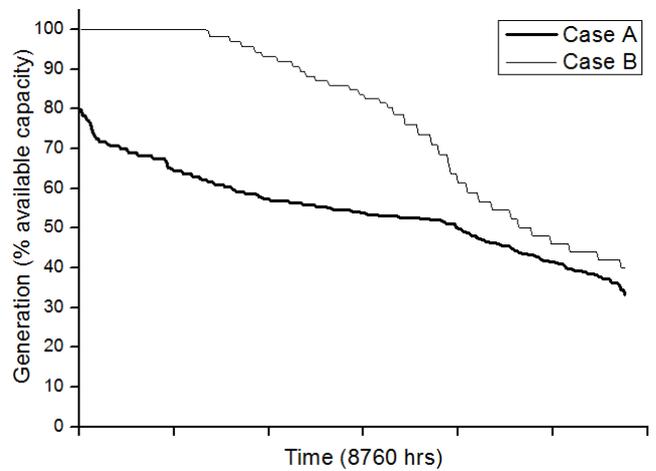

Fig. 4.b Scenario-II

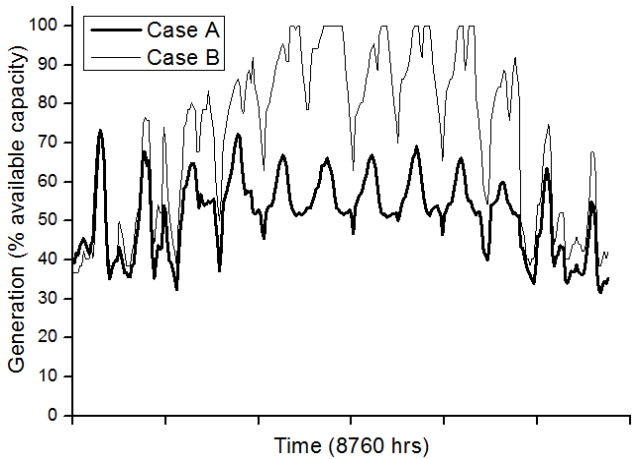

Fig. 3.c Scenario-III

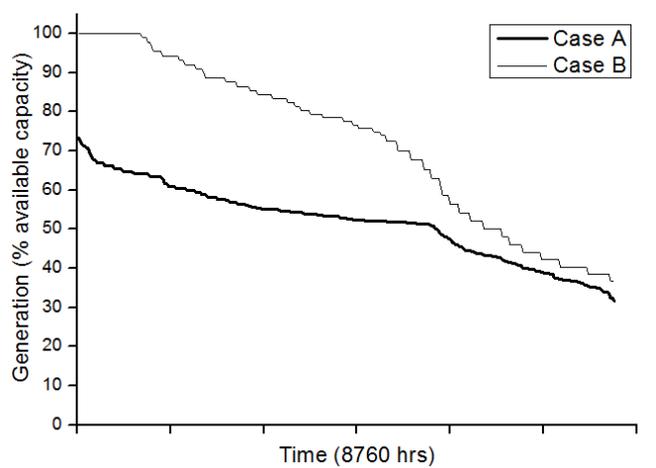

Fig. 4.c Scenario-III

Fig. 3 Annual generation in percentage of available capacity

Fig. 4 Duration curves for the generation in percentage of available capacity cases

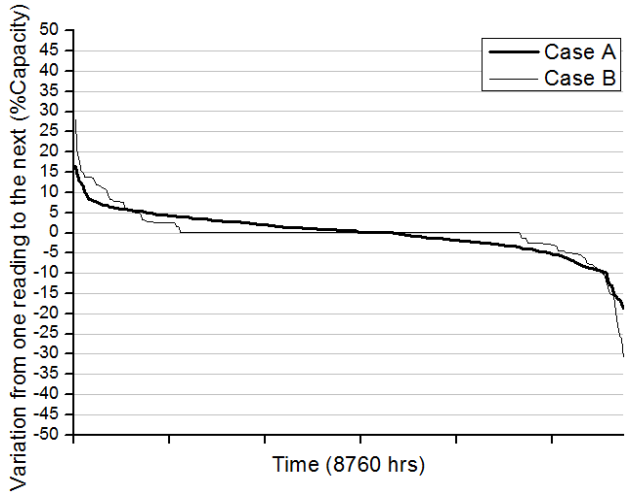

Fig. 5 a Scenario-I

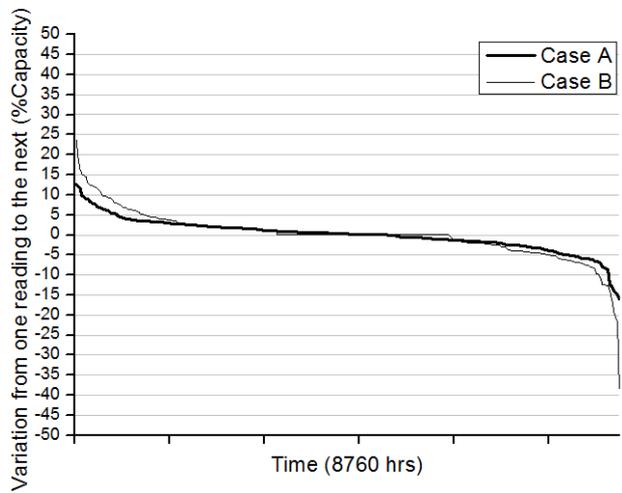

Fig. 5 b Scenario-II

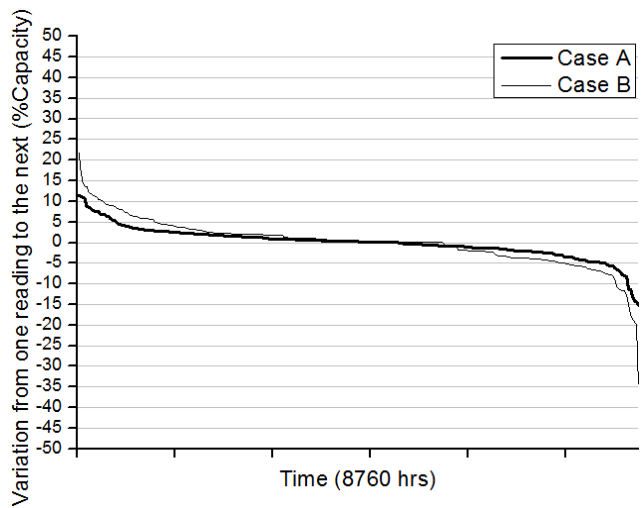

Fig. 5 c Scenario-III

Fig. 5 Duration curves of monthly variations, as a percentage of the installed capacity

Thus, to rely on a wind produced power, a conventional generation with the same value has to be included in the system to ensure its reliability. There are a number of methods to calculate the wind capacity values [23]:

✓ Probabilistic methods, based on a system load duration curve. This method is best suited for system planners. The basic principles underlying probabilistic methods to assess the capacity credit of a wind plant are standard techniques normally used to evaluate the reliability of a power system. Probabilistic methods can be subdivided into two groups; Analytical and Simulation methods

✓ The Effective Load Carrying Capability (ELCC) is considered to be the preferred metric for evaluating the capacity value of a wind farm. The ELCC is typically calculated using a power system reliability model and the conventional ELCC calculation requires substantial reliability modeling. It is an iterative process and is computationally intensive. The Non-iterative method deals with minimal reliability modeling and is computationally less intensive than the conventional approach. Since calculation of ELCC involves considerable data and computational effort, some regions such as the Pennsylvania-New Jersey-Maryland (PJM) and Southwest Power Pool (SPP) use approximate methods to calculate wind capacity credit.

✓ Approximate methods are useful when the ELCC cannot be determined due to data or other limitations. In the PJM method, the capacity credit is calculated based on the wind generator's capacity factor during the hours from 3 PM to 7 PM, from June 1 through August 31. The capacity credit is a rolling three year average, with the most recent year's data replacing the oldest year data. On the otherhand, the New York Independent System Operator (NYISO) determines wind capacity credit using the wind generator's capacity factor between 2 PM and 6 PM from June through Aug and 4 PM through 8 PM from December through February. The California Public Utilities Commission (CPUC) uses a three year rolling average of the monthly average wind energy generation between 12

PM and 6 PM for the months of May through September.

Due to the limitation of data availability, the PJM method was implemented to calculate the capacity credit for EL-Zayt site- near to Zafarana. The data used is the grid peak hours in Egypt from 5 to 8 PM, during the period from May $1^{st.}$ through August $31^{st}$. The capacity credit is averaged for three years.

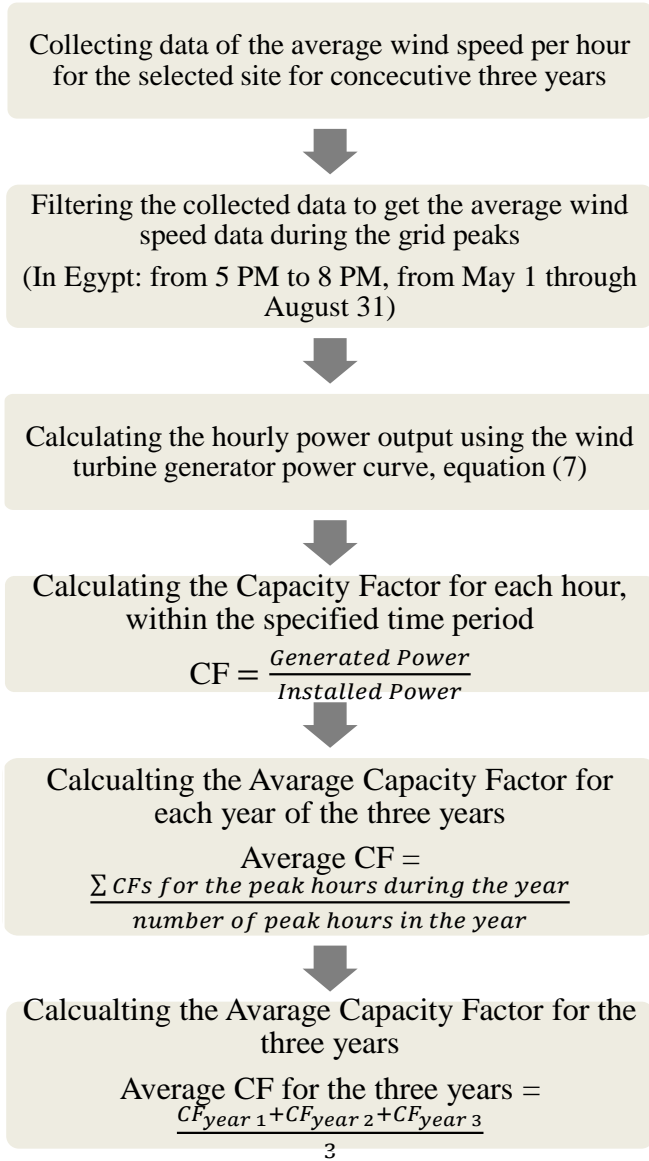

Fig. 6 Flow Chart for Capacity Credit calculations using the PJM method

Figure 6 shows the flow chart of calculating capacity credit using the PJM method. In this paper, the PJM method was applied on El-Zayt site using actual wind data of the years 2007, 2008, and 2009. Table 4 shows that the capacity credit of El-Zayt site is very high. This is due to the good correlation between wind speeds in this site and the grid peak times. In the meanwhile, the PJM method was applied using different WT characteristics as discussed in Scenarios-I to III. Results show that as the rated and cut-out speeds increase, the capacity credit of the WF decreases as less power is generated from wind.

Table 4
Capacity Credit (in %) for El Zayat Site for WTs used in Various Scenarios

|  | Scenario-I | Scenario-II | Scenario-III |
|---|---|---|---|
| 2007 | 64.76 | 54.81 | 49.78 |
| 2008 | 75.88 | 64.99 | 59.45 |
| 2009 | 80.6 | 71.14 | 65.57 |
| Average | 73.75 | 63.65 | 58.27 |

**6. Strategic Planning (Case Study in Egypt)**

As mentioned earlier, thermal plants with the same value as the WFs' capacity credit have to be installed to satisfy the system's required reliability. In this paper, it is proposed to set the timing and sizing of NPP to be installed in Egypt to cover the new WFs capacity credit. In practical applications, NPP will be running as base load, while, the same value of already existing fast speed thermal plants will be used to cover the wind fluctuations and balance the demand.

The previously mentioned cases (Case-A and B) and scenarios (Scenarios-I to III) are carried out to install a total of 3000 MW WFs. It has to be mentioned that around 500 MW is already installed at Zafarana, i.e. 2500 MW will be newly installed which will distributed as follws:
- In Case-A, 1500 MW will be installed in Zafarana and another 1500 MW in El Dabaa region starting 2018.
- In Case-B, the whole 3000 MW will be in the Zafarana/Red Sea region

Table 5 shows the coordination of the NPP timing and sizing with respect to WFs installations. For both cases A and B, as the rated and cut-out speeds increase, i.e. from Scenario-I to III, the output generated power decreases. This leads to reducing the capacity credit of the WF and in return the timing and sizing value of the NPP to be installed. Table 5 shows the values of NPP that cover the WF capacity credit for each scenario under study. The gray cells indicate that the timing and the rest of the NPP sizing can be delayed giving room for decision makers to evaluate the existing NPP installations. These results can play a significant role in the planning phase of selecting the WTs characteristics. Even though, Case-A (which represents the WF geographical distribution), generates less wind power, it gives other technical, economical benefits as well as provides more room for feedback from decision markers. That is shown by more gray cells than those reported by Case-B.

Table 5
NPP Timing and Sizing (MW) Selection Based on WFs MW Installations

| Year | | Till 2012 | 2016/ 2017 | 2017/ 2018 | 2018/ 2019 | 2019/ 2020 | 2020/ 2021 |
|---|---|---|---|---|---|---|---|
| Case-A | | | | | | | |
| WF | Z | 500 | 500 | 500 | | | |
| | D | | | | 500 | 500 | 500 |
| NPP-Scenario | I | | | | 2000 | | |
| | II | | | | 1000 | | |
| | III | | | | 1000 | | |
| Case-B | | | | | | | |
| WF | | 500 | 500 | 500 | 500 | 500 | 500 |
| NPP-Scenario | I | | | | 2000 | | 1000 |
| | II | | | | 2000 | | |
| | III | | | | 1000 | 1000 | |

Z: Zafarana, D: El Dabaa

## 7. Conclusions and Recommendations

In this paper, the benefits of coupling WF with NPP are evaluated. Different aspects of this coupling are studied. The characteristics of the site that can accommodate both NPP and WF are mentioned, and the existence of sites in Egypt that can accommodate coupling is verified. The geographical distribution of WFs installation in Egypt is examined on two practical sites (Cases-A and B); Zafarana and El Dabaa Sites. The WFs capacity credit is calculated using the PJM method using practical three years hourly wind speed data at the Red Sea region. Three scenarios are carried out for each case under study to examine the impact of selecting various WTs rated and cut-out speeds. Finally, a proposed methodology is reported using the obtained capacity credit values to help decision makers to evaluate and adapt the strategic expansion plan for NPP timing and sizing. The studies carried out in this paper showed that the positive impact of coupling NPP and WF and the WF geographical distribution is significant for wide area countries. However, in countries with narrow areas, these benefits can be achieved if there is an interconnection with neighbour countries that in return widens the coupling and geographical areas.